\UseRawInputEncoding

\documentclass[aps,showpacs,twocolumn,pra,superscriptaddress]{revtex4}

\usepackage[tbtags]{amsmath}
\usepackage{amssymb}
\usepackage{amsfonts}
\usepackage{bmpsize}
\usepackage{mathtools}
\usepackage[colorlinks=true,citecolor=blue,urlcolor=blue]{hyperref}

\hypersetup{colorlinks,linkcolor=blue,filecolor=blue,urlcolor=blue,citecolor=blue,}
\usepackage{multirow}

\begin{document}

\title{Full Network nonlocality sharing in extended scenario via Optimal Weak Measurements}

\author{Zinuo Cai}
\affiliation{Key Laboratory of Low-Dimensional Quantum Structures and Quantum Control of Ministry of Education, Key Laboratory for Matter
Microstructure and Function of Hunan Province, Department of Physics and Synergetic Innovation Center for Quantum Effects and
Applications, Hunan Normal University, Changsha 410081, China}
\author{Changliang Ren}\thanks{Corresponding author: renchangliang@hunnu.edu.cn}
\affiliation{Key Laboratory of Low-Dimensional Quantum Structures and Quantum Control of Ministry of Education, Key Laboratory for Matter Microstructure and Function of Hunan Province, Department of Physics and Synergetic Innovation Center for Quantum Effects and Applications, Hunan Normal University, Changsha 410081, China}

\begin{abstract}
Quantum networks, which can exceed the framework of standard bell theorem, flourish the investigation of quantum nonlocality further. Recently, a concept of full quantum network nonlocality (FNN) which is stronger than network nonlocality, has been defined and can be witnessed by Kerstjens-Gisin-Tavakoli (KGT) inequalities [\href{https://link.aps.org/doi/10.1103/PhysRevLett.128.010403}{Phys. Rev. Lett. 128 (2022)}]. In this letter, we explored the recycling of FNN as quantum resources by analyzing the FNN sharing between different combinations of observers. The FNN sharing in extended bilocal scenario via weak measurements has been completely discussed. According to the different motivations of the observer-Charlie$ _1$, two types of possible FNN sharing, passive FNN sharing and active FNN sharing, can be investigated by checking the simultaneous violation of KGT inequalities between Alice-Bob-Charlie$ _1$ and Alice-Bob-Charlie$ _2$. Our results show that passive FNN sharing is impossible while active FNN sharing can be achieved by proper measurements, which indicate that FNN sharing requires more cooperation by the intermediate observers compared with Bell nonlocal sharing and network nonlocal sharing.
\end{abstract}


\maketitle

\section{Introduction}

Bell inequality admits local-hidden-variable (LHV) models, but can be violated in the quantum world \cite{PhysicsPhysiqueFizika.1.195,PhysRevLett.23.880,PhysRevLett.65.1838,PhysRevA.46.5375,belinskiui1993interference,PhysRevLett.88.210401,2014Bell}. This ingenious idea makes it possible to test the conflict between quantum mechanics and classical intuition \cite{schrodinger1936indeterminism} through experiments for the first time. Along with the experimental observation \cite{PhysRevLett.28.938,PhysRevLett.49.1804,PhysRevLett.64.2495,PhysRevLett.81.3563,PhysRevLett.81.5039,mair2001entanglement} of the violation of Bell inequality, it gave birth to a new discipline-quantum information science \cite{1984Quantum,bennett2000quantum,bouwmeester2000physics,jaeger2007quantum}. In the following decades, Bell nonlocality which reveals from the violation of Bell inequality, has been extended from various perspectives \cite{PhysRevD.35.3066,RevModPhys.65.803,PhysRevLett.71.4287,PhysRevLett.88.170405,PhysRevLett.89.060401,gisin2002local,PhysRevLett.101.050403,Wood_2015}, such as extending Bell-type inequalities to more complex multi-partite and high-dimensional system \cite{PhysRevD.35.3066,RevModPhys.65.803,PhysRevLett.71.4287,PhysRevLett.88.170405,PhysRevLett.89.060401}. A complete background on Bell inequalities has been summarized in \cite{2014Bell}. Nevertheless, in the blooming of the investigation of Bell nonlocality, there exist two obscure characteristics. One is that, each local particle is usually measured only once in each round of Bell tests. Another is, no matter how many-party scenarios, all particles emit from the same source.

With the depth of exploration of bell nonlocality, physicists gradually realized these hidden limitations of the original scenarios, and began to break through them in order to deeply understand quantum nonlocality in a broad perspective. On one side, in contrast to the standard Bell scenario, the improved scenario that the same qubit is measured sequentially by multiple different observers in each round of bell test has been discussed recently. In 2015, Silva et al. \cite{PhysRevLett.114.250401} originally demonstrated that Bell nonlocality can be shared among more than two observers using weak measurements. In their scenario, a 2-qubit entangled state is distributed to three observers Alice, Bob$ _1$, and Bob$ _2$, in which Alice receives the first qubit and the two Bobs receive the second qubit. Alice carries out a strong measurement on her received qubit, while Bob$ _1$ measures the received qubit weakly and then passes it to Bob$ _2$, who then performs strong measurements independently and is unaware of Alice's result. Surprisingly, they exhibited a counterintuitive results in this scenario that a simultaneous violation of Clauser-Horne-Shimony-Holt (CHSH) inequalities between Alice-Bob$ _1$ and Alice-Bob$ _2$ is possible, which is counterintuitive. Since then, a series of investigations of nonlocal sharing has been reported both in theories \cite{PhysRevLett.114.250401,mal2016sharing,bera2018witnessing,PhysRevA.100.052121,PhysRevA.100.062130,PhysRevA.103.052207,PhysRevA.104.L060201,PhysRevA.105.032211} and experiments \cite{Schiavon_2017,hu2018observation,PhysRevA.102.032220,PhysRevLett.125.080403,PhysRevResearch.2.033205,Anwer_2021}. Wherein some of typical extensions include active and passive nonlocality sharing \cite{PhysRevA.100.052121}, bilateral nonlocal sharing\cite{bera2018witnessing,PhysRevA.105.032211}, quantum contextuality \cite{PhysRevA.100.062130,Anwer_2021}, quantum communication \cite{PhysRevLett.125.080403,PhysRevResearch.2.033205}, quantum steering sharing \cite{PhysRevA.103.052207}, and so on. The exploration of nonlocal sharing, which is also called recycling nonlocality via sequential measurements, provides a special perspective for deeply understanding quantum correlation.

On the other side, following the development of quantum network, understanding the quantum correlations in networks becomes more and more important. In contrast of the previous research on quantum correlation, quantum correlations in networks involve many independent sources rather than one, which is a significant feature of quantum networks. A series of studies on network nonlocality that may transcend Bell nonlocality has been reported \cite{PhysRevLett.104.170401,fritz2012polyhedral,PhysRevA.85.032119,PhysRevA.96.020304,PhysRevLett.121.140501,PhysRevA.90.062109,Saunders_2017,Tavakoli_2021,Tavakoli_2022} very recently. The bilocal scenario \cite{PhysRevLett.104.170401,PhysRevA.85.032119,PhysRevA.96.020304,PhysRevLett.121.140501}, which corresponds to the scenario underlying entanglement swapping experiments \cite{PhysRevLett.71.4287,ukowski1995EntanglingPR,PhysRevA.60.194}, is the simplest quantum network scenario, in which two independent sources share entangled pairs with three observers. Branciard et.al \cite{Saunders_2017} showed that network nonlocality can be exhibited by the violation of a network Bell inequality-Branciard-Rosset-Gisin-Pironio inequality, which opens the door to studying network nonlocality. In 2022, Kerstjens et.al \cite{Pozas_Kerstjens_2022} explained the concept of full network nonlocality, and showed that it is stronger than standard network nonlocality, where BRGP inequality does not witness full network nonlocality.

Undoubtedly, it is a particularly interesting prospect to study the recycling network nonlocality. While either recycling nonlocality or network nonlocality is in the process of rapid development independently, the investigation of the recycling network nonlocality is still very limited. Recently, Hou et.al \cite{PhysRevA.105.042436} investigated network nonlocality sharing based on weak measurements in the extended bilocal scenario, and soon afterward similar work was also reported \cite{wang2022network}. In this paper, we investigate the full network nonlocality sharing phenomenon in the extended bilocal scenario. According to \cite{Pozas_Kerstjens_2022}, full network nonlocality can be witnessed via violations of KGT inequalities where two inequalities should be violated simultaneously. Firstly, we show the maximal violation of KGT inequalities in a bilocal scenario, which can achieve $\sqrt{13}\approx 3.6055$ when Alice and Charlie$ _1$ chose the proper measurements. Secondly, inspired by the investigation of Bell nonlocality sharing \cite{PhysRevA.100.052121,PhysRevA.103.052207} and network nonlocality sharing \cite{PhysRevA.105.042436}, the passive FNN sharing and active FNN sharing depending on the different motivations of the observer-Charlie$ _1$ are investigated respectively. Through analyzing the simultaneous violation of KGT inequalities between Alice-Bob-Charlie$ _1$ and Alice-Bob-Charlie$ _2$, we demonstrate that it is impossible to observe passive FNN sharing even if Charlie$ _1$ carries out proper weak measurement with the optimal pointer. However, active FNN sharing can be observed, as the simultaneous violation of KGT inequalities between Alice-Bob-Charlie$ _1$ and Alice-Bob-Charlie$ _1$ exists. In addition, the noise immunity of full network nonlocality (sharing) has also been discussed.

\begin{figure}[htbp]
	\centering
	\includegraphics[width=0.45\textwidth]{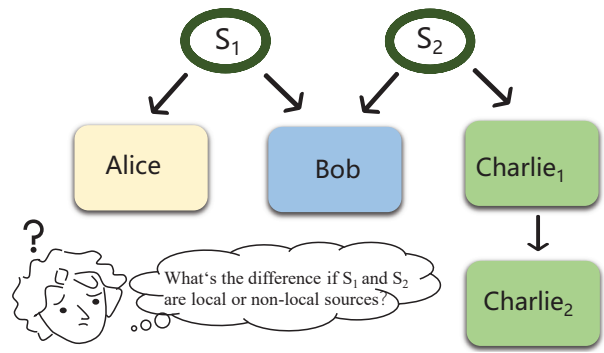}
	\caption{\small{The extended bilocal scenario, $S_1$ and $S_2$ are two sources. The arrows represent the transfer of particles. A two-qubit entangled state produced by the source $S_1$ is shared with Alice and Bob, and another entangled state produced by the source $S_2$ is shared with Bob and Charlie$ _1$, subsequently Charlie$ _1$ delivers it to Charlie$ _2$. After these observers's independent measurements, the outcome statistics is network nonlocal (NN), when it cannot be explained by any BLHV model with two different local hidden variables. The correlation of this network is fully network nonlocal (FNN) if it even cannot be explained in any model where one of the two sources is represented by a local hidden variable and another one by general no-signaling resources (NS).}} \label{f1}
\end{figure}

The structure of the paper is as follows: In Sec.\ref{full2}, we review the description of FNN in the extended bilocal scenario. Subsequently, we tightly demonstrated the maximal violation of KGT inequalities in a bilocal scenario, as is shown in Sec.\ref{full3}. Passive FNN sharing and active FNN sharing are completely analyzed in Sec.\ref{full4} and \ref{full5}. The noise resistance of active FNN sharing is discussed. We ended the paper with a conclusion in Sec.\ref{full6}.

\section{The Full network non-locality in the extended bilocal Scenario}\label{full2}

We start by recalling the description of full network non-locality in the bilocal scenario \cite{PhysRevLett.104.170401}. Quantum network links multiple remote observers using multiple independent sources. The simplest constructure, bilocal scenario, consists of two sources and three observers, as illustrated in Fig.\ref{f1}. Each source has a two-particle state and distributes the qubits to the two neighboring observers. The first source ($S_1$) sends its qubits to Alice and Bob, and the second source ($S_2$) sends its qubits to Bob and Charlie$ _1$. Without loss of generality, we assume that Alice, Bob, and Charlie$ _1$ carry out independent measurements $\hat{A}$,$\hat{B}$, and $\hat{C}$, with corresponding binary outcomes $a,b$, and $c$ respectively. As is well known, when the joint probabilities $p(a,b,c|\hat{A},\hat{B},\hat{C})$ can be always represented as
\begin{multline}
	p(a,b,c|\hat{A},\hat{B},\hat{C})=\int d\lambda _1d\lambda _2\mu _1(\lambda _1)\mu _2(\lambda _2)\\
	p(a|\hat{A},\lambda _1)p(b|\hat{B},\lambda _1,\lambda _2)p(c|\hat{C},\lambda _2)
\end{multline}
The correlation of this network is bilocal, otherwise, it is network nonlocal (NN), where $\lambda _i$ ($i\in \left \{ 1,2 \right \} $) is the local variables associated with the sources. While full network nonlocality (FNN) is strictly stronger than standard NN \cite{Pozas_Kerstjens_2022}, where FNN requires all links in a network to distribute nonlocal resources. For FNN, the joint probabilities $p(a,b,c|\hat{A},\hat{B},\hat{C})$ can not be represented as
\begin{multline}
	p(a,b,c|\hat{A},\hat{B},\hat{C})=\int d\lambda \mu (\lambda )
	p(a|\hat{A},\lambda )p(b,c|\hat{B},\hat{C},\lambda )
\end{multline}

It is not trivial to determine the FNN in quantum networks, which can not be determined by the most well-known network Bell test \cite{Saunders_2017}. Recently, Kerstjens \emph{et.al} provided a method to determine FNN based on the inflation of
networks \cite{Pozas_Kerstjens_2022}. Here we describe how to determine the FNN in an extended bilocal scenario according to \cite{Pozas_Kerstjens_2022}.
As illustrated in Fig.\ref{f1}, in contrast with the standard bilocal scenario, two observers, Charlie$ _1$ and Charlie$ _2$, will measure their shared qubits sequentially in the extended bilocal scenario. The two separate sources are $\rho _{AB}$ and $\rho _{BC}$ respectively and the initial state of the whole system can be written as,
\begin{eqnarray}
	\rho _{ABC}=\rho _{AB}\otimes\rho _{BC}.
\end{eqnarray}

Different from the measurements of the standard entanglement swapping setup, Bob performs incomplete Bell state measurement on his received qubit, denoted by $\hat{B}$, with three possible outcomes $b\in\left \{ 0,1,2 \right \} $, which corresponds to $\left | \phi^+  \right \rangle, \left | \phi^-  \right \rangle $, and the indiscernible between the two remaining Bell states $\left \{ \psi  ^+, \psi  ^- \right \} $ respectively, where $\left | \phi  ^\pm   \right \rangle=\frac{1}{\sqrt{2} } (\left | 00  \right \rangle \pm \left | 11 \right \rangle )$ and $\left | \psi  ^\pm   \right \rangle=\frac{1}{\sqrt{2} } (\left | 01  \right \rangle \pm \left | 10 \right \rangle )$. Wherein $\rho_{b}$ is the projective density matrix of the incomplete Bell state measurement with the outcome $b$. Alice (Charlie$ _m$) chose two different dichotomic measurement independently which is defined as $\hat{A}\in\{\hat{A}_{i}\}$ ( $\hat{C}_m\in\{\hat{C}_{m,j}\}$,$m\in \left \{ 1,2 \right\} $), with the corresponding outcomes $a_i\in \left \{ 0,1 \right \} $($c_{m,j}\in \left \{ 0,1 \right \} $). Wherein $\hat{A}_{i}$ represents the ith measurement of the observer Alice, $\hat{C}_{m,j}$ represents the jth measurement of the observer Charlie$ _m$. Without loss of generality, these operators can be given as,
\begin{eqnarray}\label{ACmeasure}
	&&\hat{A}_{i}= \cos \theta _{i}\sigma _x+\sin \theta _{i}\sigma _z\nonumber\\
	&&\hat{C}_{m,j}= \cos \theta _{m,j}\sigma _x+\sin \theta _{m,j}\sigma _z
\end{eqnarray}
where $\sigma _x$ and $\sigma _z$ are Pauli operators. After multiple rounds of measuring process, it is easy to obtain the joint conditional probability distribution $P(a,b,c_{1},c_{2}|\hat{A},\hat{B},\hat{C}_{1},\hat{C}_{2})$. Furthermore, the joint measurement probability distribution of three different observers, whether Alice-Bob-Charlie$ _1$ or Alice-Bob-Charlie$ _2$ can be given by the marginal constraint of the joint probability $P(a,b,c_{1},c_{2}|\hat{A},\hat{B},\hat{C}_{1},\hat{C}_{2})$. Therefore, it is possible to determine the FNN correlation between three different observers, Alice, Bob, and Charlie$ _m$.

According to \cite{Pozas_Kerstjens_2022}, the FNN can be witnessed via violations of KGT inequalities where two inequalities should be violated simultaneously,
\begin{widetext}
\begin{multline}
	\mathcal{R}_{C-NS}^{m}=
	2\left \langle A_0B_1C_{m,0} \right \rangle  -2\left \langle A_0B_1C_{m,1} \right \rangle
	+ 2\left \langle A_1B_0C_{m,0} \right \rangle  +\left \langle A_1B_0C_{m,1} \right \rangle
	-\left \langle B_0 \right \rangle  +
	\left \langle C_{m,1} \right \rangle
	[\left \langle A_1B_0 \right \rangle +\left \langle B_0C_{m,0} \right \rangle -\left \langle C_{m,0} \right \rangle ]
	\le 3
	\label{e16}\\	
	\end{multline}
\begin{equation}\label{e17}
	\begin{split}
		\mathcal{R}_{NS-C}^{m}=&2\left \langle A_0B_1C_{m,0} \right \rangle  -2\left \langle A_0B_1C_{m,1} \right \rangle
		+ \left \langle A_1B_0C_{m,0} \right \rangle
		+2\left \langle A_1B_0C_{m,1} \right \rangle
		-\left \langle B_0 \right \rangle  +
		\left \langle A_1 \right \rangle \left \langle A_1B_0 \right \rangle
		+\left \langle A_1 \right \rangle \left \langle B_0C_{m,1} \right \rangle \\&
		+ \left \langle A_1 \right \rangle \left \langle C_{m,0} \right \rangle  -
		\left \langle A_1 \right \rangle \left \langle C_{m,1} \right \rangle-
		\left \langle A_1 \right \rangle \left \langle A_1 \right \rangle
		\le 3
	\end{split}
\end{equation}
and these average value terms can be obtained by
\begin{eqnarray}\label{AVERAGE-1}
		&&\left \langle A_iB_0C_{m,j} \right \rangle =\sum_{a,c_m=0,1}(-1)^{a+c_m}[P(a,0,c_m|\hat{A}_i,\hat{B},\hat{C}_{m,j})+P(a,1,c_m|\hat{A}_i,\hat{B},\hat{C}_{m,j})-P(a,2,c_m|\hat{A}_i,\hat{B},\hat{C}_{m,j})]\\
		&&\left \langle A_iB_1C_{m,j} \right \rangle= \sum_{a,c_m=0,1}(-1)^{a+c_m}[P(a,0,c_m|\hat{A}_i,\hat{B},\hat{C}_{m,j})-P(a,1,c_m|\hat{A}_i,\hat{B},\hat{C}_{m,j})].
	\end{eqnarray}
\end{widetext}

Then, we carefully derive the required joint measurement probabilities $P(a,b,c_m|\hat{A}_i,\hat{B},\hat{C}_{m,j})$ by following the measurement process. When Bob performs incomplete BSM on the received qubits with the outcome $b$, the state of the whole system can be obtained by,
\begin{eqnarray}
	\rho _{ABC}^{b} =\left (  I\otimes \rho_{b}\otimes I \right )
	\cdot\rho_{ABC}\cdot
	\left (  I\otimes \rho_{b}\otimes I\right ) ^{\dagger }.
\end{eqnarray}
The conditional reduced system on Alice'side and Charlie'side is given by tracing out Bob's system,
\begin{eqnarray}
	\rho _{AC}^{b} =tr_{B}\left ( \rho _{ABC}^{b} \right ).
\end{eqnarray}

As has been mentioned, Alice carries out the strong measurement on her received qubit $\hat{A}\in\{\hat{A}_{i}\}$ with the outcome $a\in\{a_i\}$, the remaining system changes to,
\begin{eqnarray}
	\rho_{\hat{A}}^{a}=U_{\hat{A} }^{a}
	\rho _{AC}^{b}
	\left (U_{\hat{A} }^{a}\right )^{\dagger }
\end{eqnarray}
where $U_{\hat{A}}^{a}= {\textstyle \prod_{\hat{A}}^{a}}\otimes I$ and ${\textstyle \prod_{\hat{A}}^{a}}= \frac{I+(-1)^{a}\hat{A}}{2}$. Instead, Charlie$ _1$ and Charlie$ _2$ measure their shared qubit sequentially.
Charlie$ _1$ performs a weak measurement $\hat{C}_{1}$ on her received qubit with the quality factor $F$ and precision factor $G$, where $F$ expresses the undisturbed magnitude of quantum system after the measurement and $G$ quantifies the information gain from Charlie$ _1$'s measurement, $G,F\in[0,1]$. A more detailed description of these two parameters of weak measurement is defined in \cite{PhysRevLett.114.250401,PhysRevA.100.052121}. Here we adopt the following relation between $F$ and $G$, $F^2+G^2=1$, which implies the weak measurement process with the optimal pointer \cite{PhysRevLett.114.250401}. When Charlie$ _1$ performs a weak measurement $\hat{C}_{1}\in\{\hat{C}_{1,j}\}$ on her received qubit with the outcome $c_1\in\{c_{1,j}\}$, the remaining system changes to,
\begin{multline}
	\rho _{\hat{C}_{1}}^{c_{1}}=\frac{F}{2}\rho_{\hat{C}}^{a}
	+\frac{1+\left (-1\right )^{c_{1}}G-F}{2}
	\left [U_{\hat{C}_{1} }^{1}	\rho_{\hat{A}}^{a}
	\left (U_{\hat{C}_{1} }^{1}\right )^{\dagger }\right ]
	\\+\frac{1-\left (-1\right )^{c_{1}}G-F}{2}
	\left [U_{\hat{C}_{1} }^{0}	\rho_{\hat{A}}^{a}
	\left (U_{\hat{C}_{1} }^{0}\right )^{\dagger }\right]
\end{multline}
where $U_{\hat{C}_{m}}^{c_{m}}=I\otimes {\textstyle \prod_{\hat{C}_{m}}^{c_{m}}}$, and ${\textstyle \prod_{\hat{C}_{m}}^{c_{m}}}\in\left \{ \frac{I+(-1)^{c_{m}}\hat{C}_{m,j}}{2}  \right \}$.
Finally, Charlie$ _2$ performs a strong measurement $\hat{C}_2\in\{\hat{C}_{2,j}\}$ on her received qubit with output $c_2\in\{c_{2,j}\}$, and the reduced state can be expressed as
\begin{eqnarray}
	\rho_{\hat{C}_{2}}^{c_2}=U_{\hat{C_{2}} }^{c_{2}}
	\rho _{\hat{C}_{1}}^{c_{1}}
	\left (U_{\hat{C}_{2} }^{c_{2}}\right )^{\dagger }.
\end{eqnarray}
The measurement process of the extended bilocal scenario has been completely described. We can obtain the whole joint probability distribution from this equation,
\begin{eqnarray}
	P(a,b,c_{1},c_{2}|\hat{A},\hat{B},\hat{C}_{1},\hat{C}_{2})=Tr[\rho_{\hat{C}_{2}}^{c_2}].	
\end{eqnarray}

The marginal probability of any combination of Alice-Bob-Charlie$ _m$ $(m\in \left \{ 1,2 \right \} )$ can be obtained from the whole joint probability distribution,
\begin{multline}\label{joint-prob}
	P(a,b,c_{m}|\hat{A},\hat{B},\hat{C}_{m})=\\
	\sum_{c_{n'}}^{n\ne n'}
	P(a,b,c_{1},c_{2}|\hat{A},\hat{B},\hat{C}_{1},\hat{C}_{2}).
\end{multline}
Therefore, all average values (\ref{AVERAGE-1}) depending on different measurements of any three observers can be easily calculated from (\ref{joint-prob}). It is possible to check whether the KGT inequalities are violated or not.

\section{The maximal violation of KGT inequalities in bilocal scenario}\label{full3}

Obviously, when $G=1$, the extended bilocal scenario regresses to the standard bilocal scenario, which is the case where Charlie$ _1$ always performs strong measurements. Firstly, we analyze the optimal exhibition FNN in this case by investigating the maximal violation of KGT inequalities. Without loss of generality, we assume that the two remote sources in the scenario are singlet states, where $\rho_{AB}=\rho_{BC}=| \psi ^-  \rangle\langle \psi ^- |$. Bob performs incomplete BSM with three outcomes. According to (\ref{ACmeasure}), the direction of each dichotomic measurement can be denoted as $\{\theta _1$, $\theta _2\}$ for Alice, and $\{\theta _3$, $\theta _4\}$ for Charlie$ _1$. It is easy to obtain the left terms of KGT inequalities, $\mathcal{R}_{C-NS}^{1}$ and $\mathcal{R}_{NS-C}^{1}$, which can be given as,
\begin{equation}\label{e22}
	\begin{split}
		\mathcal{R}_{C-NS}^{1}=&\cos \theta _2(2\cos \theta _3+\cos \theta _4)+\sin \theta _1(\sin \theta _3-\sin \theta _4)\\&
		-\frac{1}{2}\sin \theta _2(2\sin \theta _3+\sin \theta _4)
	\end{split}
\end{equation}
and
\begin{equation}
	\begin{split}\label{e23}
		\mathcal{R}_{NS-C}^{1}=&\cos \theta _2(\cos \theta _3+2\cos \theta _4)+\sin \theta _1(\sin \theta _3-\sin \theta _4)\\
		&-\frac{1}{2}\sin \theta _2(\sin \theta _3+2\sin \theta _4)
	\end{split}
\end{equation}

Using the method of Lagrange multipliers, the maximal KGT inequalities violation, $Max\left \{ min\left \{  \mathcal{R}_{C-NS}^{1}, \mathcal{R}_{NS-C}^{1} \right \}  \right \}$, can be obtained as,
\begin{eqnarray}\label{e24}
	\mathcal{R}_{C-NS}^{1}=\mathcal{R}_{NS-C}^{1}=3\cos \theta _3+2\sin \theta _3\le \sqrt{13}
\end{eqnarray}
when $\theta _1=\frac{\pi }{2} ,\theta _2=0,\theta _3=-\theta _4$. And the maximal violation $\sqrt{13}\approx 3.6055$ can be achieved when $\theta _3=\mathrm{\arccos}[\frac{3}{\sqrt{13}}] $.
It is larger than the previous result presented in \cite{Pozas_Kerstjens_2022}, which shows the violation reaches $\frac{5}{\sqrt{2} }\approx 3.5355$ when $\theta _1=0,\theta _2=\frac{\pi }{2} ,\theta _3=-\theta _4=\frac{\pi}{4}$ respectively. A larger violation is more friendly to the experimental demonstration.

\section{Passive FNN sharing in the extended Bilocal scenario IS IMPOSSIBLE}\label{full4}

More generally, no matter what the reason, if Charlie$ _1$ actually performs a weak measurement rather than a strong measurement, the initial quantum system is not fully changed by the measurement process, and the corresponding initial correlation information is not completely destroyed. It allows us to explore the FNN sharing between different combinations of observations. Taking the simplest model as an example, Charlie$ _1$ chooses two different dichotomic observables randomly, and then unbiasedly delivers the measured qubit to the Charlie$ _2$, where the "unbiasedly" means the received qubits by Charlie$ _2$ equally come from any of these two different measurements by Charlie$ _1$. In addition, Charlie$ _2$ carries out strong measurements on the received qubits independently, and any communication is forbidden for all observers. In the extended bilocal scenario, we can investigate the FNN of Alice-Bob-Charlie$ _m$ ($m\in \left \{ 1,2 \right \} $), or there both.

Assumed that the direction of each dichotomic measurement can be denoted as $\{\theta _1$, $\theta _2\}$ for Alice, $\{\theta _3$, $\theta _4\}$ for Charlie$ _1$ and $\{\theta _5$, $\theta _6\}$ for Charlie$ _2$, we can obtain the KGT expressions for Alice-Bob-Charlie$ _m$ respectively according to the discussion in Sec. \ref{full2}. $\mathcal{R}_{C-NS}^{1}$ and $\mathcal{R}_{NS-C}^{1}$ corresponds to Alice-Bob-Charlie$ _1$, $\mathcal{R}_{C-NS}^{2}$ and $\mathcal{R}_{NS-C}^{2}$ corresponds to Alice-Bob-Charlie$ _2$. These different KGT expressions can be given as,
\begin{widetext}
	\begin{eqnarray}\label{KTG-1}
		\mathcal{R}_{C-NS}^{1}=G( \cos \theta _2(2\cos \theta _3+\cos \theta _4)+\sin \theta _1(\sin \theta _3-\sin \theta _4)
		-\frac{1}{2}\sin \theta _2(2\sin \theta _3+\sin \theta _4))
	\end{eqnarray}
	\begin{eqnarray}\label{KTG-2}
		\mathcal{R}_{NS-C}^{1}=G(\cos \theta _2(\cos \theta _3+2\cos \theta _4)+\sin \theta _1(\sin \theta _3-\sin \theta _4)
		-\frac{1}{2}\sin \theta _2(\sin \theta _3+2\sin \theta _4) )
	\end{eqnarray}
	\begin{equation}\label{KTG-3}
		\begin{split}
			\mathcal{R}_{C-NS}^{2}=&\frac{1}{16}(4 (1 + F) \cos[\theta _2 - \theta_5]
			-6 (-1 + F) \cos[\theta_2 + 2 \theta_3 - \theta_5]
			-6 (-1 + F) \cos[\theta_2 + 2 \theta_4 - \theta_5]
			+12 (1 + F) \cos[\theta_2 + \theta_5]\\&
			-2 (-1 + F) (\cos[\theta_2 - 2 \theta_3 + \theta_5] +\cos[\theta_2 - 2 \theta_4 + \theta_5])
			+2 (1 + F) \cos[\theta_2 - \theta_6]
			-3 (-1 + F) \cos[\theta_2 + 2 \theta_3 - \theta_6]\\&
			-3 (-1 + F) \cos[\theta_2 + 2 \theta_4 - \theta_6]
			+ 6 (1 + F) \cos[\theta_2 + \theta_6]
			-(-1 + F2) (\cos[\theta_2 - 2 \theta_3 + \theta_6]
			+\cos[\theta_2 - 2 \theta_4 + \theta_6])\\&
			-4 (-1 + F) \sin\theta_1 \sin[2 \theta_3 - \theta_5]
			-4 (-1 + F) \sin\theta_1 \sin[2 \theta_4 - \theta_5]
			+8 (1 + F) \sin\theta_1 \sin\theta_5\\&
			+4 (-1 + F) \sin\theta_1 \sin[2 \theta_3 - \theta_6]
			+ 4 (-1 + F) \sin\theta_1 \sin[2 \theta_4 - \theta_6]
			- 8 (1 + F) \sin\theta_1 \sin\theta_6	
			)
		\end{split}
	\end{equation}
	\begin{equation}\label{KTG-4}
		\begin{split}
			\mathcal{R}_{NS-C}^{2}=&\frac{1}{16}(2 (1 + F) \cos[\theta _2 - \theta_5]
			-3 (-1 + F) \cos[\theta_2 + 2 \theta_3 - \theta_5]
			-3 (-1 + F) \cos[\theta_2 + 2 \theta_4 - \theta_5]
			+6 (1 + F) \cos[\theta_2 + \theta_5]\\&
			- (-1 + F) (\cos[\theta_2 - 2 \theta_3 + \theta_5] +\cos[\theta_2 - 2 \theta_4 + \theta_5])
			+4 (1 + F) \cos[\theta_2 - \theta_6]
			-6 (-1 + F) \cos[\theta_2 + 2 \theta_3 - \theta_6]\\&
			-6 (-1 + F) \cos[\theta_2 + 2 \theta_4 - \theta_6]
			+ 12 (1 + F) \cos[\theta_2 + \theta_6]
			-2(-1 + F) (\cos[\theta_2 - 2 \theta_3 + \theta_6]
			+\cos[\theta_2 - 2 \theta_4 + \theta_6])\\&
			-4 (-1 + F) \sin\theta_1 \sin[2 \theta_3 - \theta_5]
			-4 (-1 + F) \sin\theta_1 \sin[2 \theta_4 - \theta_5]
			+8 (1 + F) \sin\theta_1 \sin\theta_5\\&
			+4 (-1 + F) \sin\theta_1 \sin[2 \theta_3 - \theta_6]
			+ 4 (-1 + F) \sin\theta_1 \sin[2 \theta_4 - \theta_6]
			- 8 (1 + F) \sin\theta_1 \sin\theta_6
			)
		\end{split}
	\end{equation}
\end{widetext}
where F and G are the quality factor and precision factor of weak measurements for Charlie$ _1$.

We can investigate the FNN sharing by analyzing the simultaneous violation of these KGT inequalities (\ref{KTG-1}-\ref{KTG-4}) between Alice-Bob-Charlie$ _1$ and Alice-Bob-Charlie$ _2$. Inspired by the deep understanding of standard Bell nonlocality sharing \cite{PhysRevA.100.052121}, the FNN sharing in the extended bilocal scenario may exist in two different types based on the motivation of Charlie$ _1$, passive and active sharing \cite{PhysRevA.100.052121} respectively. Passive FNN sharing implies that Charlie$ _1$ has no conscious thought of FNN sharing with subsequent observers, but only wants to achieve a maximal KGT violation of Alice-Bob-Charlie$ _1$. While active FNN sharing means Charlie$ _1$ wants to help the subsequent observers possess FNN as much as possible on the premise of ensuring that he has already possessed.

We discuss the possibility of passive FNN sharing at first. The pointer type of the weak measurement adopts the optimal, that is $F^2+G^2=1$, which always gives the optimal tradeoff between the information gain and disturbance \cite{PhysRevLett.114.250401}. Since Charlie$ _1$ only wants to observe the maximal violation of KGT inequalities of Alice-Bob-Charlie$ _1$ in the case of fixed $G$. According to the above result in Sec.\ref{full3}, the maximal KGT violation will be reached when $ \theta _1=\frac{\pi }{2},\theta _2=0,\theta _3=-\theta _4,\theta _3=\arccos[\frac{3}{\sqrt{13} } ]$, where $\mathcal{R}_{1}=\mathcal{R}_{C-NS}^{1}=\mathcal{R}_{NS-C}^{1}=\sqrt{13}G$. Under these optimal settings, the KGT expressions for Alice-Bob-Charlie$ _1$ are limited by the accuracy factor G.

Subsequently, Charlie$ _1$ delivers the measured qubit to Charlie$ _2$. Similarly, Charlie$ _2$ will choose the optimal measurements to achieve the maximal violation of KGT inequalities of Alice-Bob-Charlie$ _2$ under the constraint of the optimal measurement settings of Charlie$ _1$. It is easy to obtain that the maximal violation of KGT inequalities of Alice-Bob-Charlie$ _2$,  $\mathcal{R}_{2}=Max\left \{ min\left \{\mathcal{R}_{C-NS}^{2},\mathcal{R}_{NS-C}^{2} \right \}  \right \}$ will achieve when $\theta _5=-\theta _6$. The KGT inequalities of Alice-Bob-Charlie$ _2$ can be expressed as,
\begin{eqnarray}\label{KTG-5}
	\mathcal{R}_{C-NS}^{2} = \mathcal{R}_{NS-C}^{2}=K \cos\theta _5 +T \sin\theta _5\le\sqrt{K^{2} +T^{2}} \nonumber\\
\end{eqnarray}
where $K=\frac{3}{2}(1+F-(-1+F)\cos 2\theta _3)$, $T=1+F+(-1+F)\cos 2\theta _3$ and $\theta _3=\arccos[\frac{3}{\sqrt{13} } ]$. The bound of (\ref{KTG-5}) reaches when $\theta _5=\arccos[\frac{K}{\sqrt{K^2+T^2} } ]$.

\begin{figure}[htbp]
	\centering
	\includegraphics[width=0.4\textwidth]{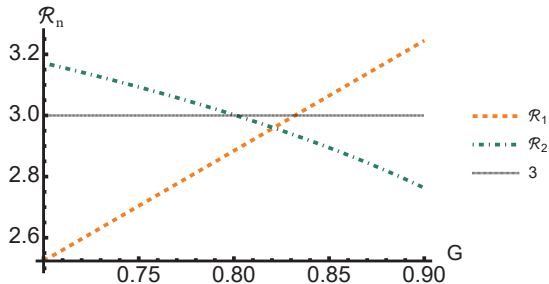}
	\caption{\small{Plot of KGT expressions $\mathcal{R}_{1}$ (orange dashed line) and $\mathcal{R}_{2}$ (green dot-dashed line) as a function of the precision factor $G$ of Alice-Bob-Charlie$ _1$ and Alice-Bob-Charlie$ _2$, $G\in [0.7,0.9]$. This scenario shows a case that Charlie$ _1$ has no conscious thought of FNN sharing with subsequent observers, but only wants to achieve a maximal KGT violation of Alice-Bob-Charlie$ _1$. $\mathcal{R}_{1}$ and $\mathcal{R}_{2}$ cannot simultaneously exceed 3. }} \label{f2}
\end{figure}

Using relations between the quality factor $F$ and the precision factor $G$, $G^2+F^2=1$, $\mathcal{R}_{1}=\sqrt{13}G$ and $\mathcal{R}_{2}=\sqrt{K^{2} +T^{2}}$ are only depend on the precision factor $G$.
The passive FNN sharing can only be observed when both $\mathcal{R}_{1}$ and $\mathcal{R}_{2}$ exceed 3 simultaneously. Unfortunately, it is easy to obtain that $\mathcal{R}_{1}$ exceed 3 in the range of $G\in [0.8321,1]$, while $\mathcal{R}_{2}$ exceed 3 in the range of $G\in [0,0.8014]$. As illustrated in Fig.\ref{f2}, the maximal value of $Max\left \{ min\left \{\mathcal{R}_{1},\mathcal{R}_{2} \right \}  \right \}$ is $2.9596$, when $G=0.8208$. The simultaneous violation of KGT inequalities between Alice-Bob-Charlie$ _1$ and Alice-Bob-Charlie$ _2$ is impossible, the passive FNN sharing phenomenon does not exist. This property of FNN sharing is different from that of Bell nonlocal sharing \cite{PhysRevA.100.052121} or network nonlocal sharing \cite{PhysRevA.105.042436}.

\section{The active FNN sharing in the extended Bilocal scenario}\label{full5}
As opposed to passive sharing, if Charlie$ _1$ is willing to help Charlie$ _2$ to exhibit FNN as much as possible under the condition that the KGT inequalities from Alice-Bob-Charlie$ _1$ can be guaranteed to be violated, it is denoted as active FNN sharing. To exhibit active FNN sharing, it is necessary to solve the question, $Max\left \{ min\left \{ \mathcal{R}_{C-NS}^{1},\mathcal{R}_{NS-C}^{1}, \mathcal{R}_{C-NS}^{2},\mathcal{R}_{NS-C}^{2} \right \}  \right \}$. Using the extreme value solution method, we obtain that the solution should satisfy $\theta _1=\frac{\pi }{2} ,\theta _2=0,\theta _3=-\theta _4,\theta _5=-\theta _6$.

\begin{figure}[htbp]
	\centering
	\includegraphics[width=0.4\textwidth]{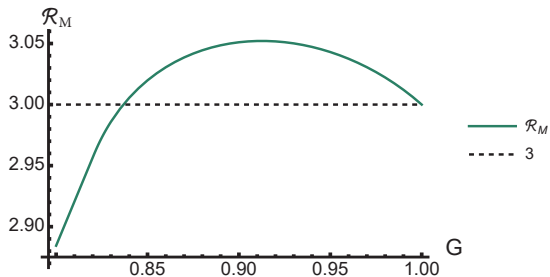}
	\caption{\small{Plot of $\mathcal{R}_{M}$ (green solid line) as a function of the precision factor $G$, $G\in [0.8,1]$. $\mathcal{R}_{M}=Max\left \{ min\left \{ \mathcal{R}_{1},\mathcal{R}_{2} \right \}  \right \}$. This scenario shows a case that Charlie$ _1$ wants to help the subsequent observers possess FNN as much as possible on the premise of ensuring that he has already possessed.}} \label{f3}
\end{figure}

\begin{figure}[htbp]
	\centering
	\includegraphics[width=0.4\textwidth]{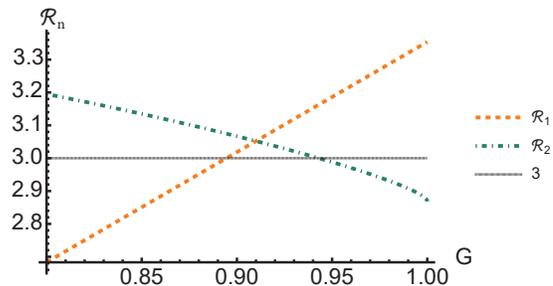}
	\caption{\small{Plot of KGT expressions $\mathcal{R}_{1}$ (orange dashed line) and $\mathcal{R}_{2}$ (green dot-dashed line) as a function of the precision factor $G$ of Alice-Bob-Charlie$ _1$ and Alice-Bob-Charlie$ _2$, $G\in [0.8,1]$. $\mathcal{R}_{1}$ and $\mathcal{R}_{2}$ exceed 3 in a range of $G\in (0.8945,0.9431)$ which means that they can violate the KGT inequalities at the same time.}} \label{f4}
\end{figure}

Taking these condition into $\mathcal{R}_{C-NS}^{1}$, $\mathcal{R}_{NS-C}^{1}$, $\mathcal{R}_{C-NS}^{2}$, and $\mathcal{R}_{NS-C}^{2}$, we can obtain that  $\mathcal{R}_{C-NS}^{1}=\mathcal{R}_{NS-C}^{1}=\mathcal{R}_{1}$, $\mathcal{R}_{C-NS}^{2}=\mathcal{R}_{NS-C}^{2}=\mathcal{R}_{2}$, where
\begin{eqnarray}\label{31}
	&&\mathcal{R}_{1}=3G\cos \theta _3+2G\sin \theta _3\\
	&&\mathcal{R}_{2}=K \cos\theta _5 +T \sin\theta _5\le \sqrt{K^{2}+T^{2}  }
\end{eqnarray}
where $K=\frac{3}{2}(1+F-(-1+F)\cos 2\theta _3)$, $T=1+F+(-1+F)\cos 2\theta _3$. Therefore, $\mathcal{R}_{1}$ is determined by $G$ and $\theta _3$, while $\mathcal{R}_{2}$ is determined by $\theta _3$ and $\theta _5$, and $G$ owing to $F^2+G^2=1$. Using the method of Lagrange multipliers, we can obtain the analytical expression of $\mathcal{R}_{M}=Max\left \{ min\left \{ \mathcal{R}_{1},\mathcal{R}_{2} \right \}  \right \}$. But the analytical expression is too complex and too long to show here. Nevertheless, $\mathcal{R}_{M}$ as a function of $G$ can be drawn clearly as illustrated by Fig.\ref{f3}. $\mathcal{R}_{1}\equiv\mathcal{R}_{2}$, when $\mathcal{R}_{M}$ reaches. The simultaneous violation of KGT inequalities between Alice-Bob-Charlie$ _1$ and Alice-Bob-Charlie$ _2$ exists in the range of $G\in(0.84,1)$, which means the active FNN sharing can be demonstrated using weak measurements with near-maximum strength. The maximal value of $\mathcal{R}_{M}$ can reaches 3.0521, when $G=0.91$, $\theta _3=0.2122$ and $\theta _5=0.2929$. As we fixed $\theta _3=0.2122$ and $\theta _5=0.2929$, a sub-optimal simultaneous violation of KGT inequalities can be observed in the range of $G\in (0.8945,0.9431)$, where $\mathcal{R}_{1}\neq \mathcal{R}_{2}$ as illustrated in Fig.\ref{f4}.

In fact, the impact of noise is inevitable. It is also interesting to discuss the noise immunity of FNN or active FNN sharing in this scenario. The noise may come from many different aspects. For simplicity, we just analyze the case of imperfect sources in the network. We assume that the shared state of these sources is not a singlet state, but a Werner state $\rho(V_i)$,
\begin{eqnarray}
	\rho(V_i)=V_i(| \psi ^-  \rangle\langle \psi ^- | )+\frac{1-V_i}{4}I\otimes I
\end{eqnarray}
where $ I$ is $2\times 2$ identity. Assuming other conditions remain constant, Charlie$ _1$ performs strong measurements, the maximal KGT inequalities violation, $Max\left \{ min\left \{  \mathcal{R}_{C-NS}^{1}, \mathcal{R}_{NS-C}^{1} \right \}  \right \}$, can be obtained that $\mathcal{R}_{C-NS}^{1}=\mathcal{R}_{NS-C}^{1}=3V_1V_2\cos \theta _3+2V_1V_2\sin \theta _3\le \sqrt{13} V_1V_2$. It is easy to find the violations of the KGT inequalities of Alice$ _1$-Bob-Charlie$ _1$ when $V\ge 0.912$, where $V=\sqrt{V_1V_2} $. Hence, the FNN can be observed if $V$ is no less than $0.912$.

Similarly, the noise immunity of the simultaneous violation of KGT inequalities between Alice-Bob-Charlie$ _1$ and Alice-Bob-Charlie$ _2$ can be discussed. For two Werner states as resources,
\begin{eqnarray}
	&&\mathcal{R}_1=V_1V_2(3G\cos \theta _3+2G\sin \theta _3)\nonumber\\
	&&\mathcal{R}_2=V_1V_2\sqrt{K^{2} +T^{2}}.
\end{eqnarray}
Supposed that $V=\sqrt{V_1V_2}$, we can observe the simultaneous violation of the KGT inequalities between Alice$ _1$-Bob-Charlie$ _1$ and Alice$ _1$-Bob-Charlie$ _2$ when $V\ge 0.9914$. Obviously, the active FNN sharing is too fragile to robust noise in the extended bilocal scenario.

\section{Conclusion}\label{full6}

We have deeply investigated the characterization of FNN in an extended bilocal scenario, where only one-sided sequential measurements were carried out by two observers. Initially, we exhibited the optimal FNN by the maximal
violation of KGT inequality of Alice-Bob-Charlie$ _1$, when the scenario regresses to the standard bilocal. It is shown that the maximal violation of KGT inequality can achieve $\sqrt{13}\approx 3.6055$ when $\theta _1=0,\theta _2=\frac{\pi }{2} ,\theta _3=-\theta _4=\mathrm{\arccos}[\frac{3}{\sqrt{13}}]$, which is larger than the previous result $\frac{5}{\sqrt{2} }\approx 3.5355$ \cite{Pozas_Kerstjens_2022} with $\theta _1=0,\theta _2=\frac{\pi }{2} ,\theta _3=-\theta _4=\frac{\pi}{4}$. The greater the violation reaches, the easier the experiment carries out. Furthermore, the recycling of FNN as quantum resources by analyzing the FNN sharing between different combinations of observers in the extended bilocal scenario has been explored. According to the different motivations of Charlie$ _1$ in their measurement process, we have completely discussed the possibility of two typical FNN sharing, passive FNN sharing and active FNN sharing respectively. Our results clearly demonstrated that the passive FNN sharing is impossible, as the maximal simultaneous achievement of KGT values $\mathcal{R}_{1}$ and $\mathcal{R}_{2}$ for Alice-Bob-Charlie$ _1$ and Alice-Bob-Charlie$ _2$ is $2.9596$, which means that the KGT inequalities cannot be simultaneously violated by these different combinations of observers Alice-Bob-Charlie$ _m$. On the contrary, when Charlie$ _1$ wants to help the subsequent observers possess FNN as much as possible on the premise of ensuring that he has already possessed, the simultaneous violation of KGT inequalities between Alice-Bob-Charlie$ _1$ and Alice-Bob-Charlie$ _2$ can be observed in the range of $G\in(0.84,1)$, which indicates the active FNN sharing exists. The simultaneous maximal violation of KGT inequalities between Alice-Bob-Charlie$ _1$ and Alice-Bob-Charlie$_2$
is 3.0521, when $G=0.91$, $\theta _3=0.2122$ and $\theta _5=0.2929$.  These results indicate that, compared with Bell nonlocal sharing and network nonlocal sharing, FNN sharing requires more cooperation by the intermediate observers. Moreover, we also discussed the noise immunity of FNN or active FNN sharing in this scenario.

\section{Acknowledgment}

C.R. was supported by the National Natural Science Foundation of China (Grant No. 12075245), the Natural Science Foundation of Hunan Province (2021JJ10033), Xiaoxiang Scholars Programme of Hunan Normal university.


\bibliographystyle{apsrev4-1}
\bibliography{ref}

\end{document}